\documentclass[11pt,a4paper]{article}

 \usepackage[english]{babel}
 \usepackage[T2A]{fontenc}
 \usepackage[cp1251]{inputenc}
 \usepackage{amsmath}
\usepackage{graphicx}
\usepackage{amssymb}
\usepackage{color}
\usepackage{amsfonts}
\usepackage{wrapfig}
\usepackage{caption}

\newcommand{\non}{\nonumber \\}

\newcommand{\be}{\begin{equation}}
\newcommand{\ee}{\end{equation}}
\newcommand{\bea}{\begin{eqnarray}}
\newcommand{\eea}{\end{eqnarray}}

\newcommand{\lp}{\left (}
\newcommand{\rp}{\right )}

\newcommand{\lbr}{\left \{}

\newcommand{\rd}{\right .}

\newcommand{\vl}{{\bf l}}

\newcommand{\cM}{{\frak M}}

\begin{document}

\binoppenalty=10000
\relpenalty=10000

\begin{center}
\textbf{\Large{PHASE BEHAVIOR OF A CELL MODEL WITH CURIE-WEISS INTERACTION}}
\end{center}

\vspace{0.3cm}

\begin{center}
M.P.~Kozlovskii and O.A.~Dobush\footnote{e-mail: dobush@icmp.lviv.ua} 
\end{center}

\begin{center}
Institute for Condensed Matter Physics of the National Academy of Sciences of
Ukraine \\ 1, Svientsitskii Str., 79011 Lviv, Ukraine
\end{center}

 \vspace{0.5cm}

\small The object of this study is a cell model with Curie-Weiss interaction potential. We have already proved~\cite{ref1,ref2} the possibility of a mathematically rigorous transition from a continuous system of interacting particles to such a model and made an accurate calculation of its grand partition function.
In the present research, we derive an explicit accurate equation of state of the cell model with Curie-Weiss potential. It turned out that this model has a sequence of first-order phase transitions at temperatures below the critical $ T_c $. We analyzed the mechanism of these transitions based on the behavior of the chemical potential as a function of density. Thus we found the points of phase coexistence without going beyond the microscopic approach. We also proposed a mathematically strict definition of the critical temperature as a function of the microscopic parameters of the model. We determined the parameters of the critical points and plotted phase diagrams which cover the area of the first three phase transitions in the sequence. 

\vspace{0.5cm}

PACs: 51.30.+i, 64.60.fd

Keywords: cell model,  Curie-Weiss interaction, equation of state, first-order phase transitions, phase diagrams

\normalsize

\renewcommand{\theequation}{\arabic{section}.\arabic{equation}}
\section{Introduction}
\setcounter{equation}{0}

We have recently proposed a kind of cell model describing the phenomenon of a first-order phase transition within the grand canonical ensemble without using any phenomenological assumptions. In the case of realistic interaction potentials, it has an approximate solution, which, however, is not related to expansions in density, activity or other small parameters. In particular, we derived the equation of state~\cite{ref3,ref4} of the model with the Morse potential, calculated the coordinates of the critical point, and obtained the temperature dependence of the order parameter in the region of subcritical temperatures. The mathematical apparatus involved the use of the space of collective variables~\cite{ref5,ref6}. Moreover, we proved~\cite{ref1,ref2} that \emph{this model has an exact solution} in the case of the Curie-Weiss interaction.

In contrast to~\cite{ref1,ref2} in the present research, we obtained explicit expressions for pressure as a function of temperature and density of such a system, and paid attention to investigating its phase behavior. Our results cover a wide range of densities and temperatures except the critical region. This is typical of the mean-field approximation. It is the type of Curie-Weiss potential that corresponds to the idea of this approximation, namely the distance-\emph{independent} interaction between particles (the field equally affects all particles). Lattice models of a similar nature were considered by B. Lev et al.~\cite{ref12,ref13}. They calculated grand partition functions of few mean-field systems and obtained exact equations of state, noting the possibility of phase transitions in these models. However they didn't make either numerical analysis or description of phase behavior. The cell gas model was introduced in~\cite{ref14,ref15} by A. Rebenko et al. The authors found that such a model describes the phase transition, calculated its free energy, established the preservation of the phase transition if the cell volume goes to zero. In this model, the number of particles in a cell is limited - no more than one.

Systems with weak interactions of infinite range occupy a special place in the study of phase behavior. Therefore, the partition function of the one-dimensional system is calculated in~\cite{ref16}. The result is the van der Waals equation along with the Maxwell construction. According to R.~Balescu~\cite{ref17}, a true first-order phase transition occurs in this model.

Subsequent studies have broadened the class of interaction potentials and strictly proved that one can directly derive the van der Waals-Maxwell theory of phase transitions from the classic partition function. The essence of the method~\cite{ref18} is to find the upper and lower limits of the partition function $ Z $ $ (V, N, \gamma) $ and the free energy $ F (V, N, \gamma) $. Both expressions are functions of the parameter $ \gamma $, which is the argument of the interaction potential $ V (r, \gamma) $. The limit transition $ \gamma \rightarrow 0 $ allows a strict calculating of the model and detecting the jump of the order parameter using the condition of convexity of free energy. However, the determination of the critical temperature $ T_c $ as well as the establishment of the temperature dependence of the obtained results is not clarified in this paper~\cite{ref18}.

The difference between the study of the equation of state, which we cover in the present manuscript, from the exact results of other authors is to obtain an explicit form of pressure as a function of both density and temperature.  Rejecting the assumption of existing no more than one particle in a cell leads to a sequence of first-order phase transitions in the cell model with the Curie-Weiss interaction. In addition to the general results obtained in~\cite{ref1}, here we determine coordinates of the critical points and investigate the behavior of the order parameter in the region of the first three phase transitions from the sequence.

\renewcommand{\theequation}{\arabic{section}.\arabic{equation}}
\section{Model}
\setcounter{equation}{0}

The object of this study is a cell model with a mean-field interaction potential. In our recent publications~\cite{ref1,ref2}, we have proved the possibility of a mathematically rigorous transition from a continuous system of interacting particles to such a model and the exact calculation of its grand partition function. As a result, we proved that a \textit{sequence of the first-order phase transitions} exists in the cell model with the Curie-Weiss interaction. In this paper, we refer to the calculation of the grand partition function of the model, which is described in detail in~\cite{ref1,ref2}, and focus on considerable additions to the existing results of analytical calculation and quantitative analysis of the phenomenon.

The properties of any system are convenient to study, having its equation of state, in particular, obtained from the well-known to statistical physics relation
\be
PV = k_B T \ln \Xi,
\ee
where $P$ is the pressure of the system, $V$ is its total volume, $k_B$ is the Boltzmann constant, $T$ is the temperature, and $\Xi$ is the grand partition function. The computation of its equation of state reduces to calculating the grand partition function. In the case of a continuous system consisting of $ N $ interacting particles residing in the volume $ V $, it has the following form~\cite{ref8}
\be\label{1d1}
\Xi = \sum_{N=0}^\infty \frac{\mathrm z^N}{N!} \int_V \mathrm dx_1 ... \int_V \mathrm dx_N \exp\left[ -\frac{\beta}{2} \sum_{x,y\in\eta} \Psi_\Lambda (x,y)\right],
\ee
where $\mathrm z =\exp(\beta\mu)$ is the activity, $\mu$ is the chemical potential, $\beta = (k_BT)^{-1}$ is the inverse temperature, $\eta=\{x_1,...,x_N\}$, $x_i$ are the coordinates of the $i$-th particle. If we conditionally divide the volume of the system by a finite number of $ N_v $ congruent cubic cells each being of a volume $ v $, then we write the potential energy in the form~\cite{ref2} 
\be\label{1d2}
\sum_{x,y\in\eta} \Psi_\Lambda(x,y) = \sum_{\vl_1,\vl_2\in\Lambda} U_{l_1l_2} \rho_{\vl_1}(\eta) \rho_{\vl_2}(\eta),
\ee
where $\rho_\vl(\eta)$ is the occupation number of the $l$-th cell, and the set $\Lambda$ of the cell vectors $\vl_i$ is defined as
\be \label{1d3}
\Lambda = \left\{ \vl = (l_1,l_2,l_3) | l_i=c m_i; m_i=0,1,...,N_1-1;\quad N_1^3 =N_{v} \right\}.
\ee
here $c$ is the side of a cell, and $N_1$ is the number of cells along each axis. 
The interaction potential $U_{l_1 l_2}$ is
\be\label{1d4}
U_{l_1 l_2}(x,y)=\lbr
\begin{array}{ll}
	-\dfrac{g_a}{N_v}, & x\in \Delta_{l_1}, \; y\in \Delta_{l_2}, \; l_1\neq l_2; \\
	\quad g_r, & x,y\in \Delta_l
\end{array}
\rd
\ee
The inequality $g_r>g_a>0$~\cite{ref1} secure the stability of the interaction.

The occupation number $\rho_\vl(\eta)$ of the $l$-th cell $\Delta_l$ is defined by the indicator $I_{\Delta_l}(x)$
\be\label{1d5}
\rho_\vl(\eta) = \sum_{x\in\eta} I_{\Delta_l}(x), \quad I_{\Delta_l}(x) = \lbr
\begin{array}{ll}
	1, & x\in \Delta_l; \\
	0, & x\neq \Delta_l.
\end{array}
\rd
\ee
The occupation numbers meet the condition $\sum \limits_{\vl\in\Lambda}\rho_\vl(\eta)=N$, which states that there are $ N $ particles in all cells. This condition is formal and does not allow to compute the average number of particles $ \left< N \right> $. To do so, we have to calculate the grand partition function, and use the relation
\be\label{1d6}
\left< N \right> = \frac{\partial\ln\Xi}{\partial\beta\mu}.
\ee
It is the equation \eqref{1d6} that establishes the connection between the chemical potential $ \mu $ and the density $\eta$ 
\be\label{1d7}
\eta = \left< N \right> / V = \bar n/v.
\ee

To optimize the choice of variables we introduce the following one
\be\label{1d8}
f = g_r / g_a,
\ee
which is the ratio of repulsion intensity to attraction intensity, moreover, according to the stability condition $f>1$.

One more parameter of the model is the cell volume $v$. In the limit $v\rightarrow 0$ one has a transition to the continuous system of particles. This aspect is covered in~\cite{ref14,ref15}, where the authors found that the phase transition in the cell gas model is preserved when $v\rightarrow 0$. However, remember that real systems consist of particles of finite size, and therefore the volume $ v $ cannot be too small.

As we proved in~\cite{ref1,ref2}, the grand partition function \eqref{1d1} in the case of interaction \eqref{1d4} equals the single integral
\be\label{1d10}
\Xi = (2\pi p_r N_v)^{-1/2} \int^\infty_{-\infty} \mathrm d y \exp[N_v \tilde E_0(y,\mu)],
\ee
where
\bea
&&
\tilde E_0(y,\mu) = - y^2 / 2 p_a + \ln L_0(y,\mu),
\non
&&
L_m(y,\mu) = \sum_{n=0}^{\infty} \frac{v^n}{n!} n^m \exp\left(-\frac{1}{2} f p_a n^2\right) \exp(y + \beta\mu) n.
\label{1d11}
\eea
Here and hereafter we will use the notation
\be\label{1d11a}
p_r=\beta g_r, \qquad p_a=\beta g_a,
\ee
The expression \eqref{1d10} is calculated in~\cite{ref2} by the Laplace method~\cite{ref19}. See that the special functions $L_m(y,\mu)$, which we had introduced~\cite{ref3}, are functions of the chemical potential $\mu$. Under such circumstances, the equation for chemical potential is a complex nonlinear equation. Therefore, an explicit expression for $\beta\mu = f(\bar y)$ can be obtained only by numerical methods~\cite{ref1}, which were respectively used to study the equation of state. 

{\textit The objective of this work is to explicitly calculate the equation of state of the cell model and study its phase behavior.}. In the end of this section we see that one can get an explicit expression for the chemical potential by making in \eqref{1d10} substitution of variables 
\be
y + \beta \mu + \ln v= z.
\label{1d12}
\ee
As a result \eqref{1d10} takes the form
\be
\Xi = (2\pi p_r N_v)^{-1/2} \int \mathrm d z \exp[N_v E_0(z, \mu)],
\label{1d13}
\ee
where
\be\label{1d14}
E_0(z, \mu) = - (z - \beta\mu-\ln v)^2 / 2 p_a + \ln K_0(z).
\ee
In contrast to $L_m(y,\mu)$ \eqref{1d11} the special functions $K_m(z)$
\be \label{1d15}
K_m(z) = \sum_{n=0}^{\infty} \frac{1}{n!} n^m \exp\left(-f \frac{p_a}{2}n^2 \right)
\exp(z n)
\ee
are not functions of the chemical potential but depend on the parameter $g_r$ (see \eqref{1d11a}) and the temperature, since
\be\label{1d16}
f p_a = p_r.
\ee
If $f=const$, only the parameter $p_a$ determines the quantity $E_0(z,\mu)$.

Like \cite{ref2}, we obtain the solution of the grand partition function in the representation \eqref{1d13} by the Laplace method $(N_v \gg 1)$
\be\label{1d20}
\Xi = \big(-p_r E_0^{''}(\bar z,\mu)\big)^{-\frac12}
\exp[N_v E_0(\bar z, \mu)],
\ee
where $E_0^{''}(\bar z,\mu)$ is the second derivative of $E_0(z,\mu)$ from \eqref{1d14} with respect to $z$
\be\label{1d21}
E_0^{''}(\bar z,\mu) = \frak M_2(\bar z) - p_a^{-1}.
\ee
The following inequality is the condition for the existence of a real solution of \eqref{1d20}
\be\label{1d22}
E_0^{''}(\bar z,\mu)<0 \qquad \text{або} \qquad p_a^{-1} > \frak M_2(\bar z),
\ee
where
\be\label{1d23}
\frak M_2(\bar z)  = K_2(\bar z) / K_0(\bar z) - (K_1(\bar z) / K_0(\bar z))^2.
\ee
The equation for extrema of $E_0(z,\mu)$ 
\be\label{1d17}
\frac{\partial E_0(z,\mu)}{\partial\mu} = 0
\ee
gives an explicit form of chemical potential $\mu$ as a function of $\bar z$
\be\label{1d17a}
\beta\mu = \bar z - p_a \frak M_1(\bar z) - \ln v.
\ee
Here
\be\label{1d18}
\frak M_1(z) = K_1(z) / K_0(z).
\ee

\renewcommand{\theequation}{\arabic{section}.\arabic{equation}}
\section{Determining critical values of parameters}\label{Sec_crit}
\setcounter{equation}{0}

According to \eqref{1d20}, an asymptotic solution $(N_v\gg 1)$ of the grand partition function is as follows
\be
\Xi = \exp \lp N_v E_0(\bar z)\rp,
\label{3d1}
\ee
where
\be
E_0(\bar z) = \ln K_0(\bar z) - \frac{p_a}{2}(\cM_1(\bar z))^2.
\label{3d3}
\ee
\begin{figure}[h!]
	\centering \includegraphics[width=0.5\textwidth]{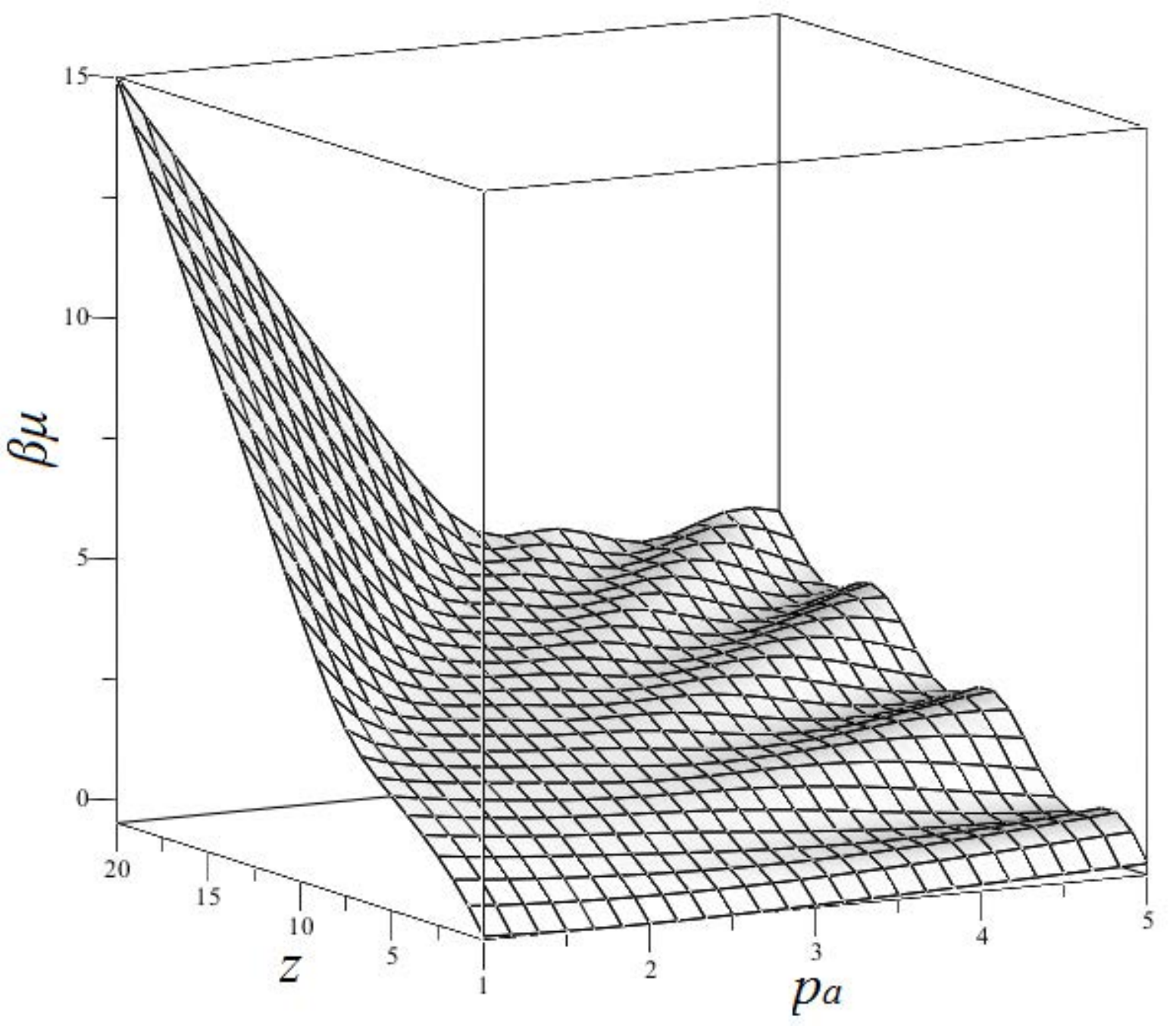}\\
	\vskip-3mm\caption{3D plot of $\beta\times \text{chemical potential}$ as a function of the variable $z$, which is linked to the density, and the parameter $p_a = \beta \times \text{attraction intencity}$. The set of the model parameters are $f=1.2$, $v=1$}\label{fig1}
\end{figure}	
The quantity $\bar z$ corresponds to $\max E_0(z)$. $z = \bar z$ only if the inequality \eqref{1d22} is satisfied. The fulfillment of \eqref{1d22} directly relates to the behavior of the chemical potential $\beta\mu$ \eqref{1d17a} as a function of $z$ (see Figure~\ref{fig1}). As we proved in~\cite{ref2}, for small values of the parameter $p_a$ (which is a part of \eqref{1d17a}) we have a mutually unambiguous correspondence between $\beta\mu$ and $\bar z$. For small $ p_a $, the quantity $ \beta \mu $ is a monotonically increasing function of its argument $ \bar z $, and stability holds for arbitrary $ z $. For large values of $ p_a (p_a> 4) $ there are certain intervals of values of $ z $, where the same value of the chemical potential corresponds to three different values of $ z $, and the function $ \beta \mu (z) $ is non-monotonic . Therefore when $p_a$ is large enough, the set of $\bar z$ values is the subset of the range of $z$.

Such behavior of the chemical potential was qualitatively revealed in~\cite{ref8} within the framework of the grand canonical ensemble. However, the lack of mathematically rigorous results for the grand partition function did not allow for quantitative analysis then. Nevertheless, it was established, that stability holds if $\mu$ is an increasing function of the density (in our case of the variable $\bar z$). Note that in~\cite{ref20} Van Hove was the first to show that within the canonical distribution, even exact calculations do not allow a loop in the equation of state of a fluid. In \cite{ref8}, it was established that within the grand canonical ensemble the derivative $ \partial \! \left <N \right> / \partial \mu $ must be strictly positive in both accurate and approximate calculations.

\begin{figure}[h!]
	\centering \includegraphics[width=0.45\textwidth]{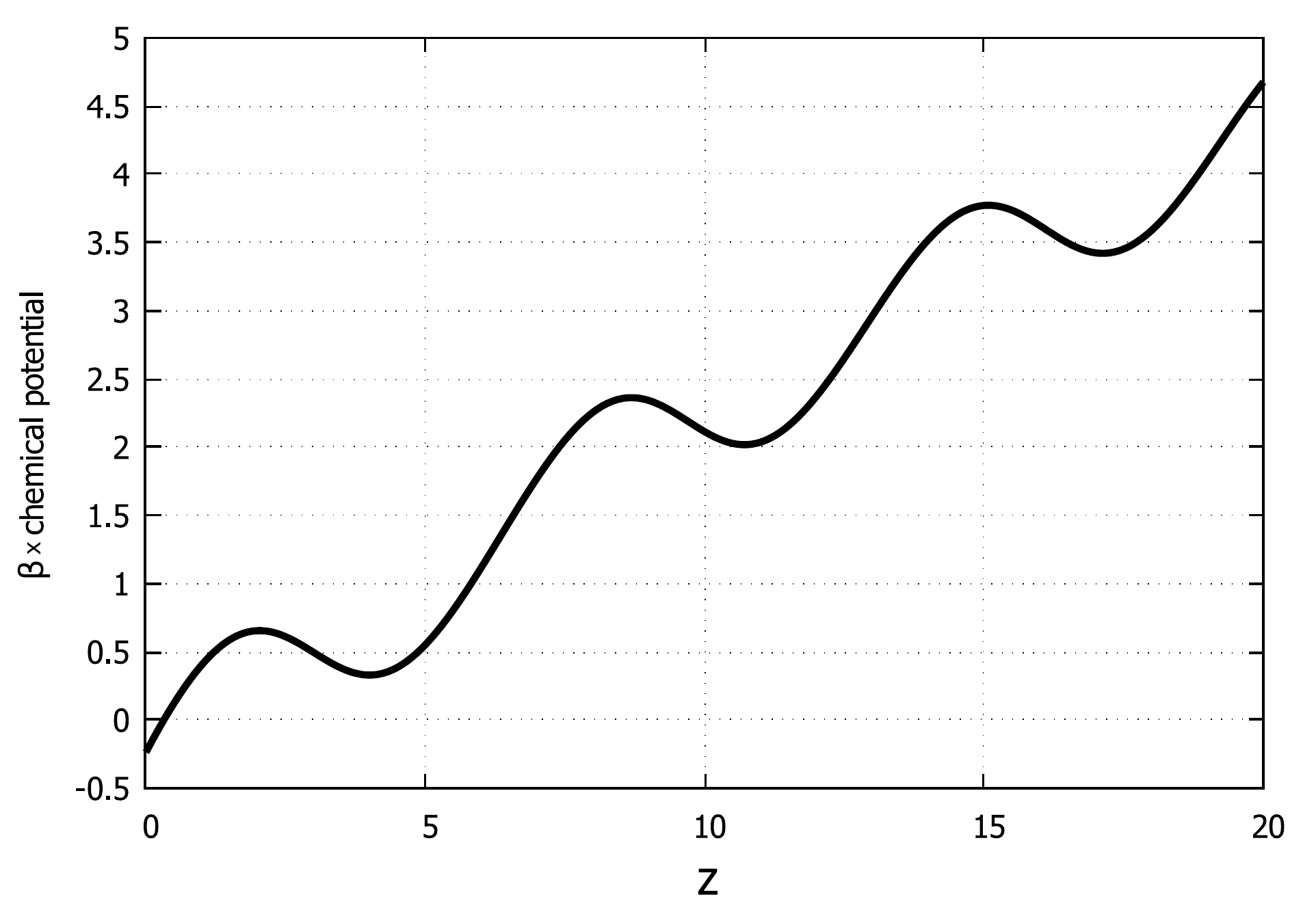}\\
	\vskip-3mm\caption{Plot of $\beta\times \text{chemical potential}$ as a function of the variable $z$ at fixed value of the parameter $p_a = 5$. The set of the model parameters are $f=1.2$, $v=1$}\label{fig2}
\end{figure}

Easy to see that for small values of the parameter $p_a=\beta g_a$ we have a monotonically increasing function $\mu(z)= \mu(\bar z)$. For large $p_a (p_a>4)$ the function $\mu(z)$ has a sequence of extrema (maxima and minima) at certain values of $z$. This fact is well seen on the plot of $\mu(z)$ at $p_a=5$ shown in Figure~\ref{fig2}. As indicated in \cite{ref8}, we clearly trace here the areas of instability where $ \partial \mu / \partial z <0 $. The chemical potential behavior points to existence of a phase coexistence, or rather to the sequence of phase transitions, in the temperature range below the critical one $T_c$, however, certain values of temperature are unknown. It is only obvious that the value of $ p_a = 5 $ is in the region $ T <T_c $.

The critical value of the parameter $p_a$ is the quantity $p_{ac}^{n}$ dividing the regimes of monotonic increasing and not monotonic behavior of the chemical potential~\cite{ref1,ref2}. The corresponding curve of the function $\mu(\bar z)$ is monotonically increasing, and in some way unique, because it has stationary inflection points. We find the latter solving the following system of equations
\be\label{3d4}
\lbr
\begin{array}{ll}
	& \cfrac{d\mu}{d z} = 0 \\
	& \cfrac{d^2\mu}{dz^2} = 0.
\end{array}
\rd
\ee
Solutions of this system is the set of certain values
\be
\{p^{(n)}_{ac}, z_c^{(n)}\}, \quad n=1,2,...,
\label{3d5}
\ee
corresponding to the critical points on the surface $\beta\mu (p_a, \bar z)$ (see Figure~\ref{fig1}). These values are represented in Table~\ref{tab1}.

\begin{table}[h!]
	\caption{Solutions of the system of equations \eqref{3d4} for critical values of the parameter $p_a$ and the variable $z$ for the first three first-order phase transitions in the sequence ($f=1.2$ and $v=1$). }	
	\label{tab1} 
	\begin{center}
		\begin{tabular}{|c|c|c|c|}
			\hline
			$(n)$ & $p^{(n)}_{ac}$ & $z_c^{(n)}$  \\ \hline
			1 & 3.928235 & 2.398348  \\ 
			2 & 3.818523 & 7.582909 \\
			3 & 3.769964 & 12.417382  \\
			\hline
		\end{tabular}
	\end{center}
\end{table}

This result is in character with the one Lebowitz and Penrose~\cite{ref18} obtained for a system with a Kac-like infinite range attraction combined with a binary repulsion. They derived the equation of state in the framework of canonical ensemble and proved the possibility of a first-order phase transition in such a system. However, their results didn't include dependence on the temperature in its explicit form. So as our parameter $p_a$ is related to temperature, but is not the temperature itself. Nevertheless, the equation of state we derived in Section 6 contains explicit pressure, density and temperature. 

Obviously, the critical values $p_{ac}^{n}$ in each of the phase transitions in the sequence differ from each other. This means that the values of critical temperatures is also different
\be
k_B T_c^{(n)} = g_a / p^{(n)}_{ac}; \quad T_c^{(n+1)} > T_c^{(n)}.
\label{3d14}
\ee
Decreasing $p^{(n)}_{ac}$ with the increase in $n$ leads to a growth of corresponding value of $T_c^{(n)}$. The specific values of the critical temperatures depend on the choice of the value of the attraction parameter $g_a$. Currently, we are more interested in the overall picture, so there is no need in assigning any certain value to $ g_a $. In addition, to simplify the notations, hereafter assume that
\be\label{3d15}
T_c^{(1)} = T_c.
\ee

\renewcommand{\theequation}{\arabic{section}.\arabic{equation}}
\section{Points of phase coexistence}\label{Sec_coex}
\setcounter{equation}{0}

\begin{figure}[h!]
	\centering \includegraphics[width=0.45\textwidth]{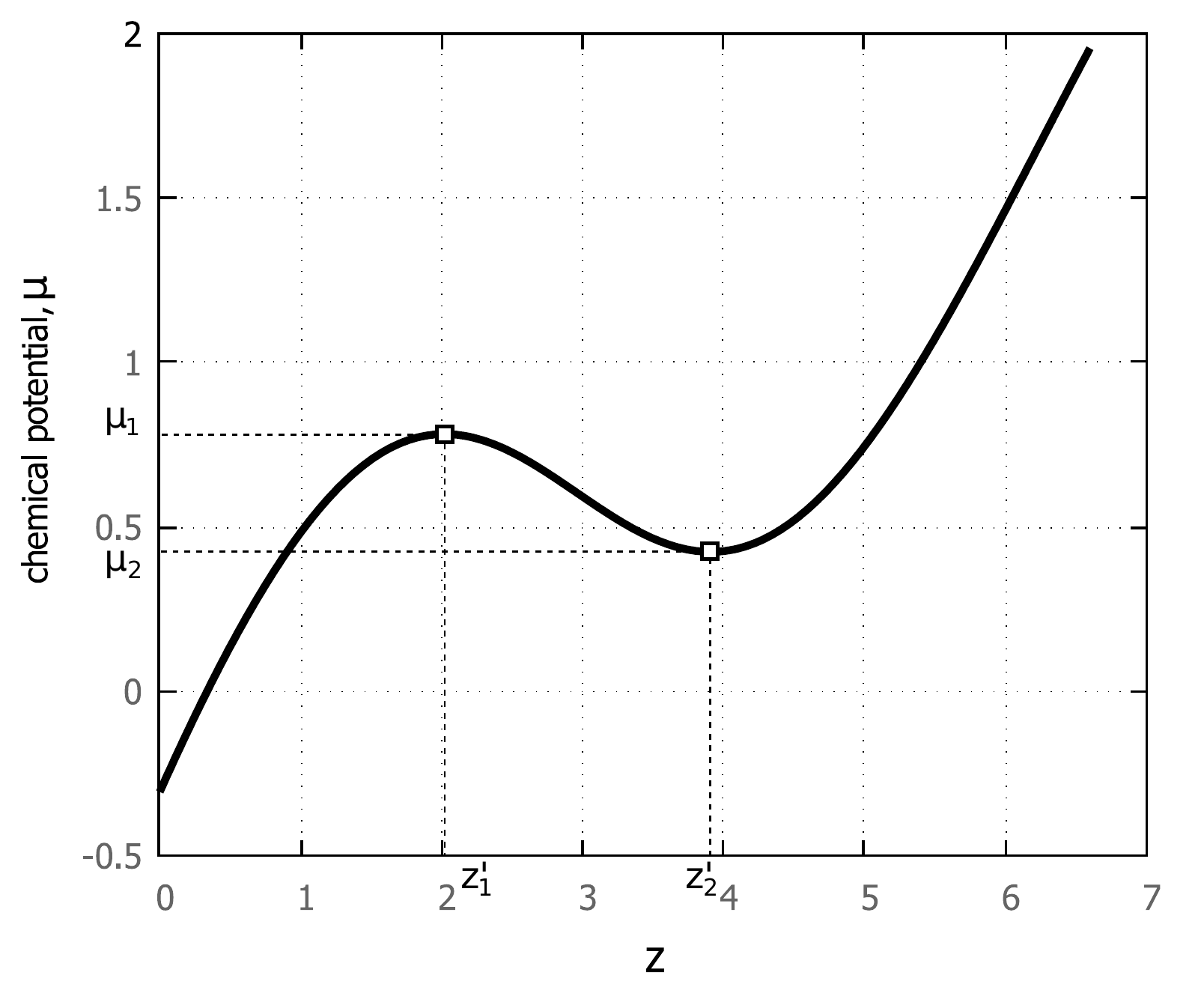}\\
	\vskip-3mm\caption{Plot of the chemical potential $\mu$ as a function of the variable $z$ at fixed value of the relative temperature $T=0.8T_c$ ($f=1.2, v=1$). Boxes are the local minima and maxima of $\mu(z)$.}\label{fig3}
\end{figure}

The next step of the behavior analysis of the investigated system is to find the points of phase coexistence at temperatures below the critical. 
At the range $p_a > p_a^{(1)}$ $(T<T_c)$ it is necessary to take into account the condition \eqref{1d22}, determining the stability of the system of interacting particles. To do this, consider in more detail the behavior of $ \mu $ (see \eqref{1d17a}) \emph {in the region of the first phase transition in the sequence}. The corresponding plot is shown in Figure~\ref{fig3}, in the case $ T = 0.8T_c $.

The function $\mu$ has a local maximum at $z=z'_1$ and a local minimum at $z=z'_2$. The chemical potential $\mu$ at $z=z'_1$ equals $\mu_1$
\be
\mu_1 = z'_1 - p_a \cM_1 (z'_1),
\label{4d7}
\ee
and at $z=z'_2$ equals $\mu_2$
\be
\mu_2 = z'_2 - p_a \cM_1 (z'_2).
\label{4d8}
\ee

In the interval $(z'_1,z'_2)$ the chemical potential decreases with increasing in the variable $ z$. Such behavior is non-physical, in this area the system is unstable.

According to \cite{ref2} \emph{solving the following system of equation we find the values $z_1$, $z_2$ of the variable $z$ determining the coexistence points}
\be\label{eq1}
\lbr
\begin{array}{ll}
	& E_0(z_1) - E_0(z_2) = 0 \\
	& \mu(z_1) - \mu(z_2) = 0.
\end{array}
\rd
\ee
\begin{figure}[h!]
	\centering \includegraphics[width=0.45\textwidth]{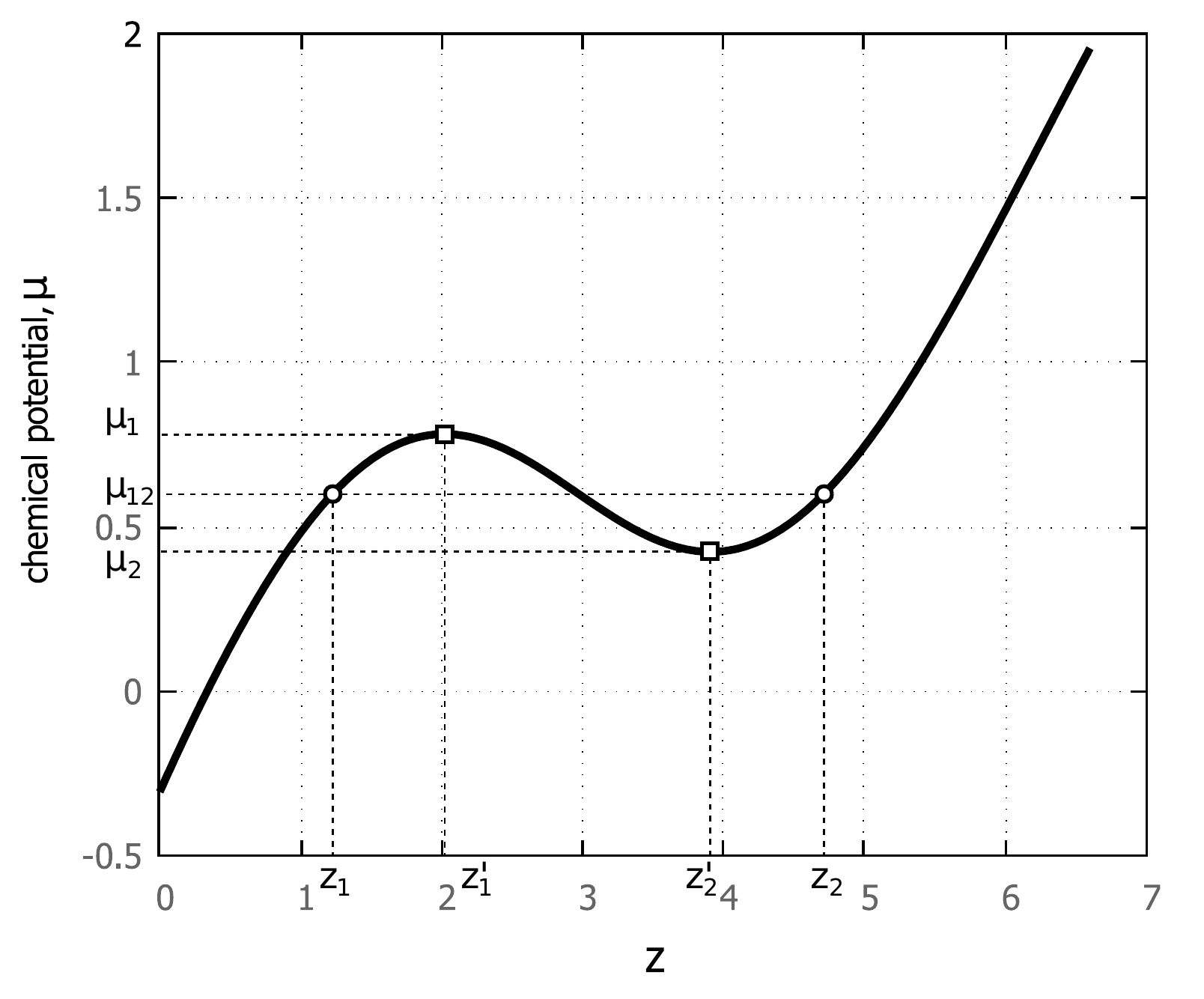}\\
	\vskip-3mm\caption{Plot of the chemical potential $\mu$ as a function of the variable $z$ at fixed value of the relative temperature $T=0.8T_c$ ($f=1.2, v=1$). Boxes are the local minima and maxima of $\mu( z)$, circles are the points of phase coexistence.}\label{fig4}
\end{figure}
\begin{figure}[h!]
	\centering \includegraphics[width=0.45\textwidth]{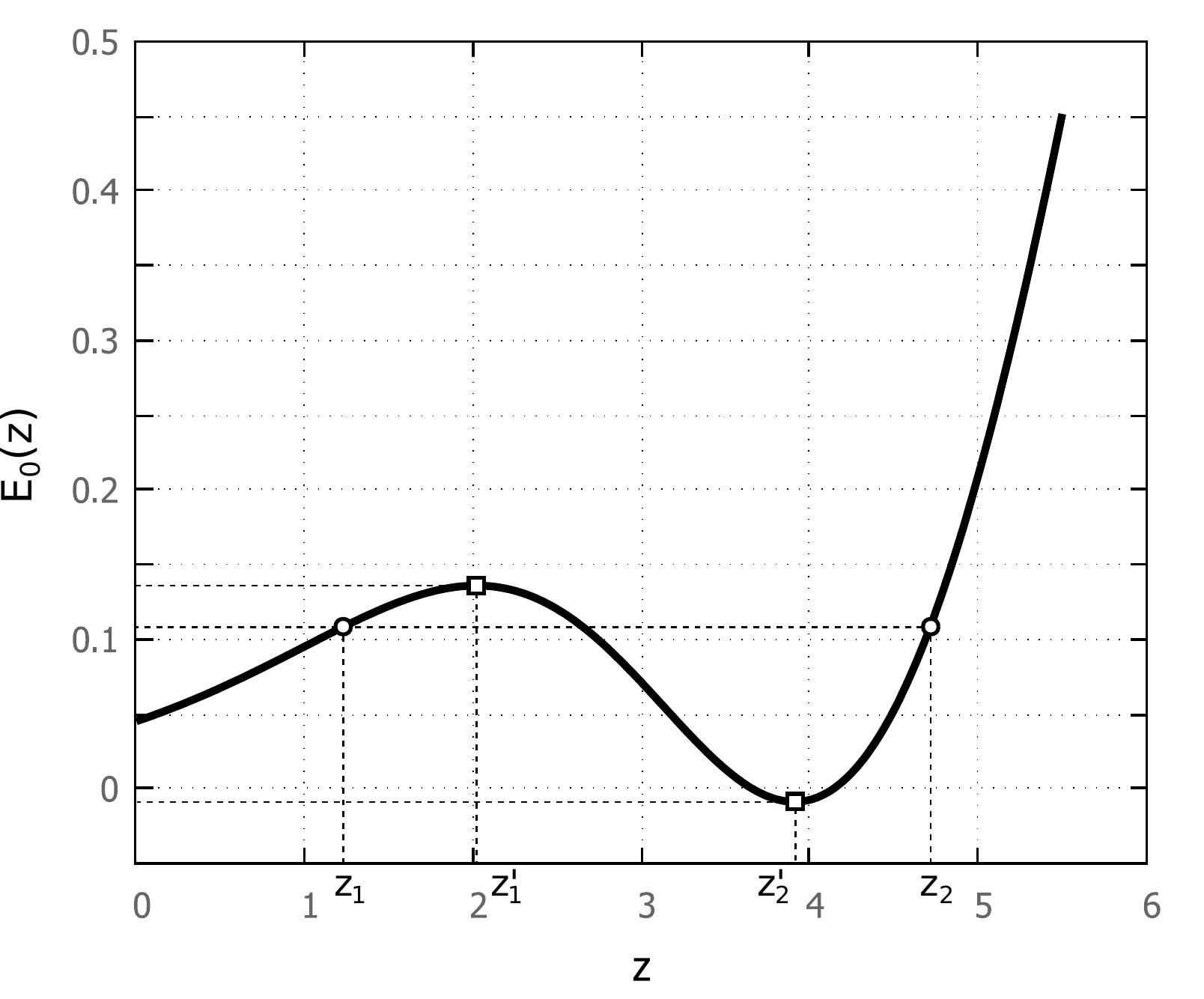}\\
	\vskip-3mm\caption{Plot of the function $E_0(z)$ at fixed value of the relative temperature $T=0.8T_c$ ($f=1.2, v=1$). Boxes are the local minima and maxima of $E_0(z)$, circles are the points of phase coexistence.}\label{fig5}
\end{figure}
Figure~\ref{fig4} and Figure~\ref{fig5} present plots of the functions $\mu(z)$ and $E_0(z)$ at the same fixed temperature $T < T_c$. Both curves have the local maximum at the point $z'_1 = 2.020156 $ and the local minimum at the point $z'_2 = 3.909327$ represented by boxes. Solutions of the system of equations \eqref{eq1} $z_1=1.230432$, $z_2=4.715610$ and corresponding values of the function $E_0(z_1) = E_0 (z_2)=0.108318$, and also the chemical potential $\mu (z_1) = \mu(z_2) = \mu_{12} = 0.602076$ represented in Figure~\ref{fig4} and Figure~\ref{fig5} by circles. For better understanding have a look at  Figure~\ref{fig6} where the function $E_0(z)$ is plotted against the chemical potential $\mu(z)$. Circle symbol represents a point on this plot corresponding to the solutions of the system of equations \eqref{eq1}. 
\begin{figure}[h!]
	\centering \includegraphics[width=0.45\textwidth]{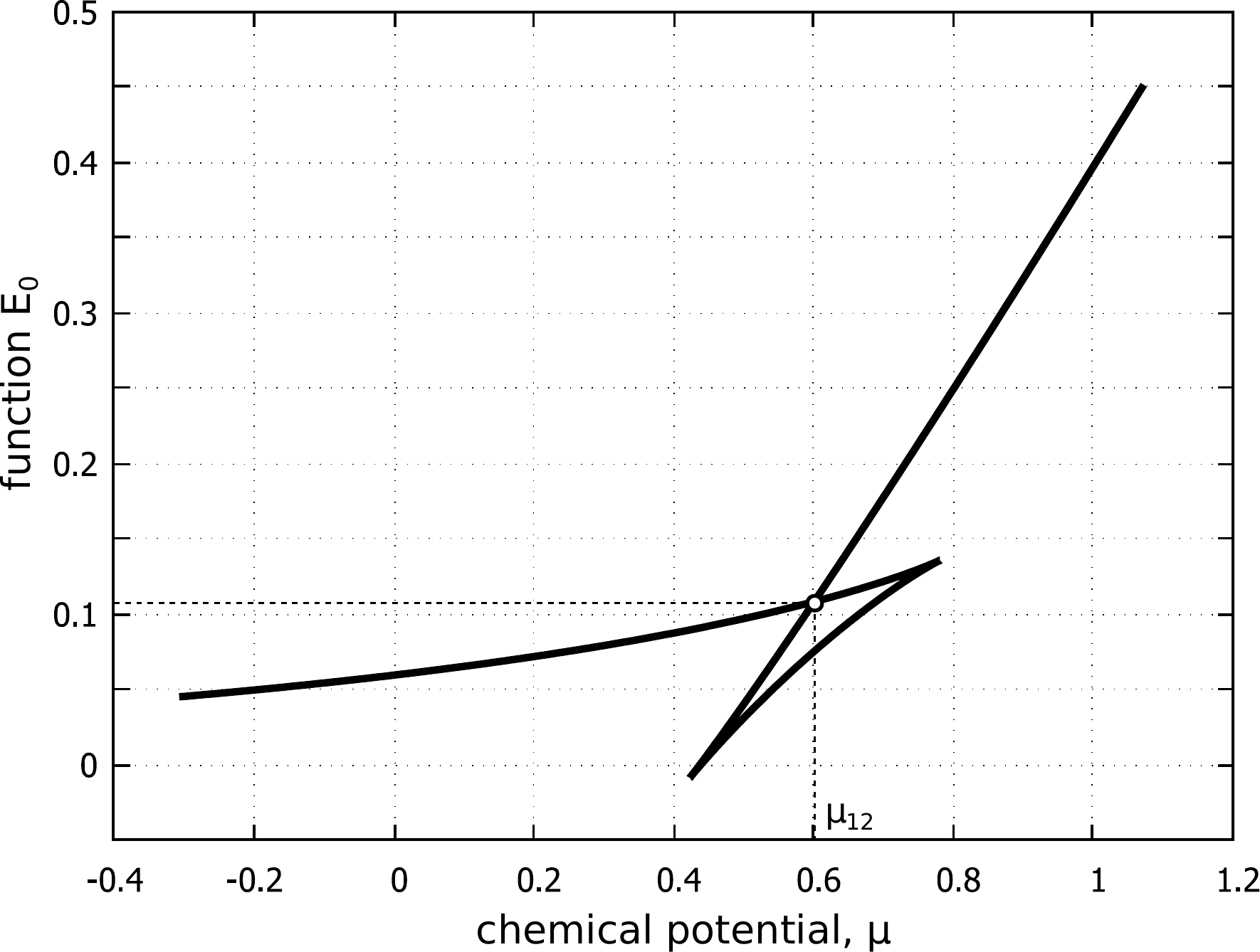}\\
	\vskip-3mm\caption{Parametric plot of $E_0(z)$ and $\mu(z)$ at fixed value of the relative temperature $T=0.8T_c$ ($f=1.2, v=1$). Circle is the point corresponding to solutions of the system of equations \eqref{eq1}.}\label{fig6}
\end{figure}

Now we see, that the function $E_0(z)$ has global maximum at $z < z_1$ and $z > z_2$. in Figure~\ref{fig6} these are the branches that go upwards from the point $(\mu_{12},E_0(z_1))$ to the right and to the left. On the intervals $(z_1,z'_1)$ and $(z'_2,z_2)$ (see Figure~\ref{fig5}) $E_0(z)$ has local maximum characterizing the metastable states of the system. According to the condition of the Laplace method, $\bar z \in (z < z_1,z > z_2)$ should be used in the region of the first phase transition to calculate the grand partition function.

\renewcommand{\theequation}{\arabic{section}.\arabic{equation}}
\section{Phase behavior of the cell model}
\setcounter{equation}{0}

\begin{figure}[h!]
	\centering \includegraphics[width=0.45\textwidth]{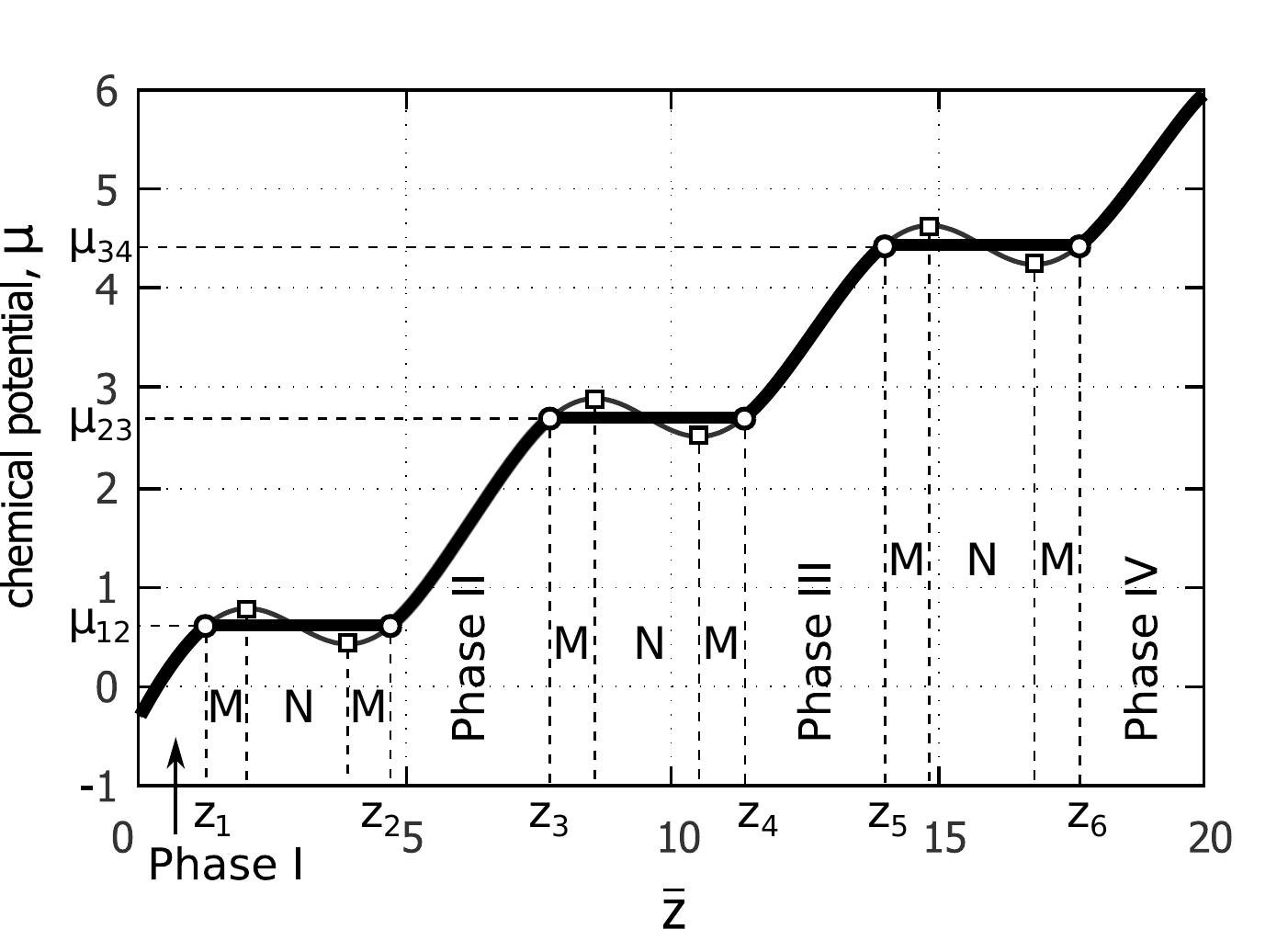}\\
	\vskip-3mm\caption{Plot of the chemical potential $\mu$ as a function of the variable $\bar z$ (bold solid line), and as a function of the variable $z$ (thin solid line) at fixed value of the relative temperature $T=0.8T_c$ ($f=1.2, v=1$). Boxes are local minima and maxima of $\mu(z)$, circles are the points of phase coexistence. Letter M denotes the region of metastable state, N --- unstable state.}\label{fig7}
\end{figure}

The procedure for determining the points of phase coexistence, described in Section~\ref{Sec_coex}, will be used to study the phase behavior of the cell model with Curie-Weiss interaction in the region of each subsequent phase transition in the sequence. Figure~\ref{fig7} presents the result of such analysis for the first three phase transitions at a fixed temperature $ T <T_c $. This figure illustrates the chemical potential as a function of both the variable $ z $ (thin line) and the variable $ \bar z $ (thick line). In Section~\ref{Sec_crit} we discussed that for $ T <T_c $ (large $ p_a $) the variable $ \bar z $ is assigned only certain values from the range of the variable $ z $. Moreover, these values have to lead to the global maxima of the function $ E_0 (z) $. We now see that this easy to understand condition allows a clear description of the nature of the first-order phase transition without going beyond the microscopic approach.

The first interval of the range of $\bar z$ is $(0,z_1)$. Here the system is in a stable state that ends in the point $(z_1,\mu_{12})$. The value of the chemical potential $ \mu_{12} $ also corresponds to another value of the variable $\bar z$, namely $z_2>z_1$. Moreover, in the interval $(z_2,z_3)$, the system resides in the next stable state different from the previous one. A jump between the values of $ \bar z $ at the same level of chemical potential implies the first-order phase transition from one stable state Phase I to the next one \--- Phase II. Consequently, the points $(z_1,\mu_{12})$ and $(z_2,\mu_{12})$ are actually points of phase coexistence. The second phase transition with phase coexistence points $(z_3,\mu_{23})$ and $(z_4,\mu_{23})$ occurs between Phase II and Phase III. The third phase transition occurs from Phase III to Phase IV at the points $(z_5,\mu_{34})$ and $(z_6,\mu_{34})$. Recall that the cell model with Curie-Weiss interaction which we study in the present research has a multitude of such first-order phase transitions. 

Under special conditions, the system remains in metastable states. in Figure~\ref{fig7} the corresponding areas are marked with letter M. The variable $ z $ takes the value between the points of coexistence and the nearest local extremes of the chemical potential (marked by boxes on the plot). In metastable areas, we see increasing chemical potential with a growth of $ z $. Such behavior is physical. However, as noted in Section~\ref{Sec_coex}, the maxima of the function $ E_0 (z) $ are not global here, so under normal conditions, such states are not achieved. Letter N presents the areas of unstable states located between the nearest points of local extrema.

\renewcommand{\theequation}{\arabic{section}.\arabic{equation}}
\section{Density of the cell model. Equation of state}
\setcounter{equation}{0}

\begin{figure}[h!]
	\centering \includegraphics[width=0.45\textwidth]{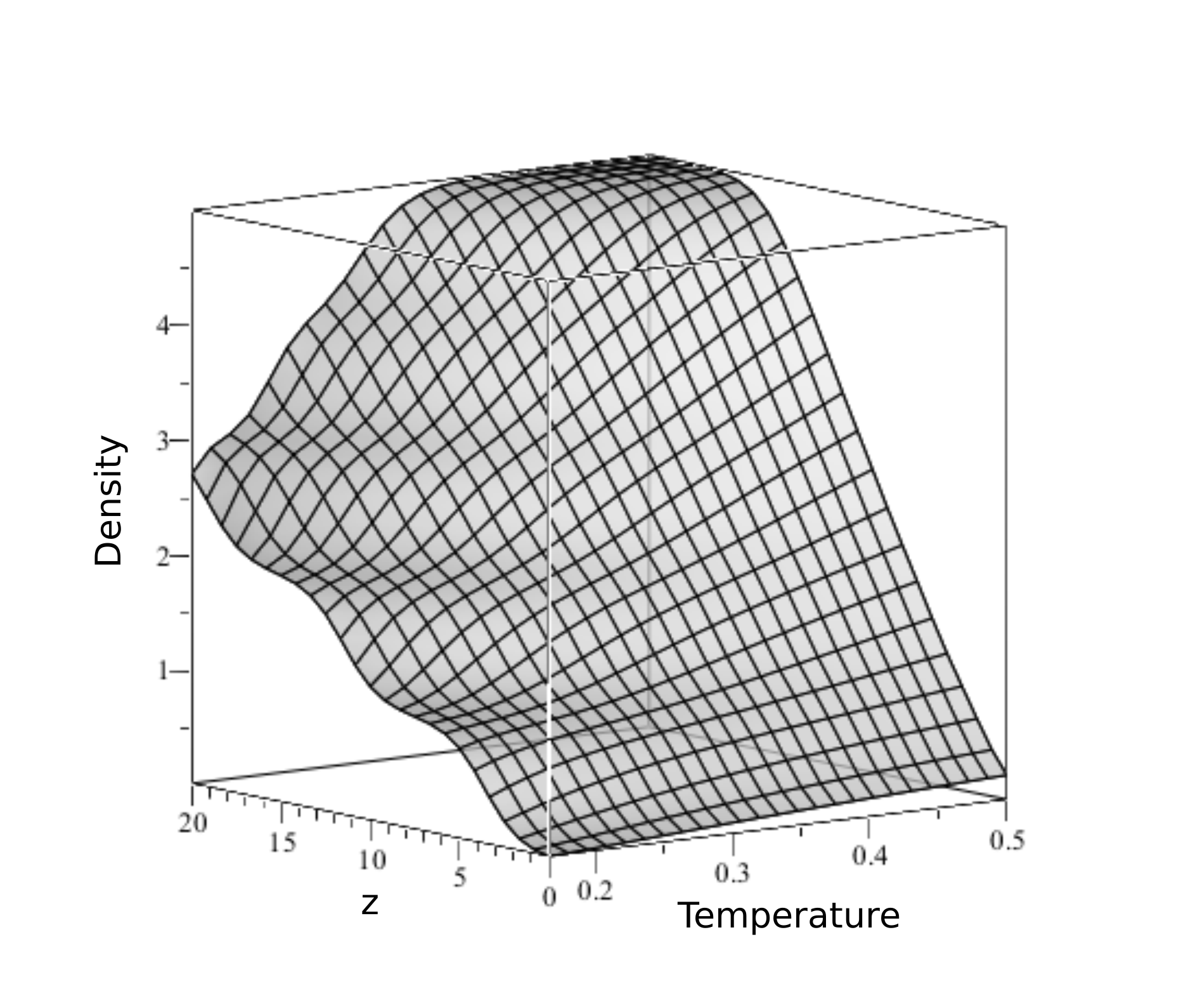}\\
	\vskip-3mm\caption{3D-plot of the density $\eta$ as a function of the variable $z$ and the temperature $const \times T$ for wide range of temperature below and above the critical one $const \times T_c = 0.254567$  ($f=1.2, v=1$).}\label{fig8}
\end{figure}

In the previous sections, we analyzed the behavior of the chemical potential as a function of the variable $ z $. However, it is sufficient to use the formulas \eqref{1d6}, \eqref{1d7} and \eqref{3d1} to obtain the expression for density
\be\label{eq3} 
\eta = \frac{1}{v}\frak M_1(z).
\ee
Therefore, the density is a monotonically increasing function of $ z $. This is clearly seen in Figure~\ref{fig8} presenting a 3D-plot of the density as a function of $ z $ and the temperature in the range $0.65 T_c <T<2 T_c$. Obviously, the equation \eqref{eq3} linking the density $\eta$ with the variable $z$ establishes mutually unambiguous relations $\eta = \eta(\bar z)$ and $\bar z = \bar z(\eta)$ both in the ranges $T>T_c$ and $T<T_c$. 

In the final stage of the research, we calculate an equation of state of the cell model in explicit form. For this purpose, use the well-known formula \eqref{1d1} connecting the pressure $P$ with the grand partition function
\be
P V = k_B T \ln \Xi \nonumber
\ee
and obtain the following
\be\label{4d5}
P v = k_B T \ln K_0(\eta) - \frac{g_a v^2}{2} \eta^2, \qquad v = V / N_v.
\ee
Note that here appears $K_0 (\eta)$ defined in \eqref{1d15} as a function of $\bar z$. However, due to the equation \eqref{eq3} one can express $\bar z$ as a function of the density $\eta$.

The equation of state \eqref{4d5} is universal for the cell model with Curie-Weiss interaction \eqref{1d4} in the sense of describing its behavior at any temperature. We used it to calculate the critical values of the thermodynamic parameters of the system given in Table~\ref{tab2}.

\begin{table}[h!]
	\caption{Parameters of critical points of the cell model with Curie-Weiss interaction in the region of first three phase transitions in the sequence ($ f = 1.2 $ and $ v = 1 $). All values are dimensionless. }	
	\label{tab2} 
	\begin{center}
		\begin{tabular}{|c|c|c|c|c|}
			\hline
			$(n)$ & $T_c^{(n)}$   & $\eta_c^{(n)}$ & $P_c^{(n)}$ & $\mu_c^{(n)}$ \\ \hline
			1     & $T_c$         & 0.5139      &  0.1977     & 0.3796        \\ 
			2     & 1.0287 $T_c$  & 1.5056      & 1.7194      & 1.7826        \\
			3     & 1.0420 $T_c$  & 2.5030      & 4.1567      & 2.8609        \\
			\hline
		\end{tabular}  
	\end{center}
\end{table}

\renewcommand{\theequation}{\arabic{section}.\arabic{equation}}
\section{Phase diagrams}
\setcounter{equation}{0}

\begin{figure}[h!]
	\centering \includegraphics[width=0.85\textwidth]{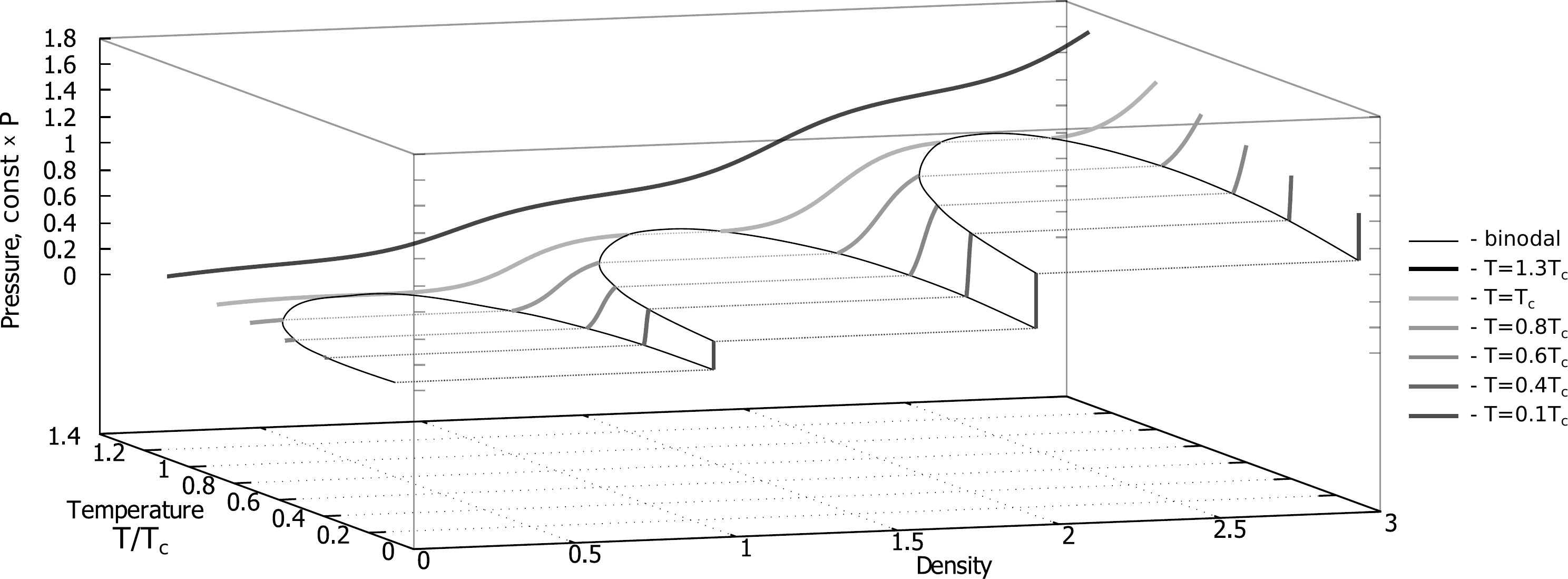}\\
	\vskip-3mm\caption{3D-plot of the isotherms, namely the pressure as a function of the density at fixed values of relative temperature $T/T_c$, in the region of first three first-order phase transitions. The case of $T\leq T_c$ is shown by different shades of solid grey lines (dotted lines present density jumps during each phase transition). At  $T> T_c$ first-order phase transitions are impossible. This is shown by a continuous, monotonically increasing isotherm (thick solid black line) in this region. Thin solid parabola-shaped lines are the phase coexistence curves (the binodals) ($f=1.2, v=1$).}\label{fig9}
\end{figure}
\begin{figure}[h!]
	\centering \includegraphics[width=0.45\textwidth]{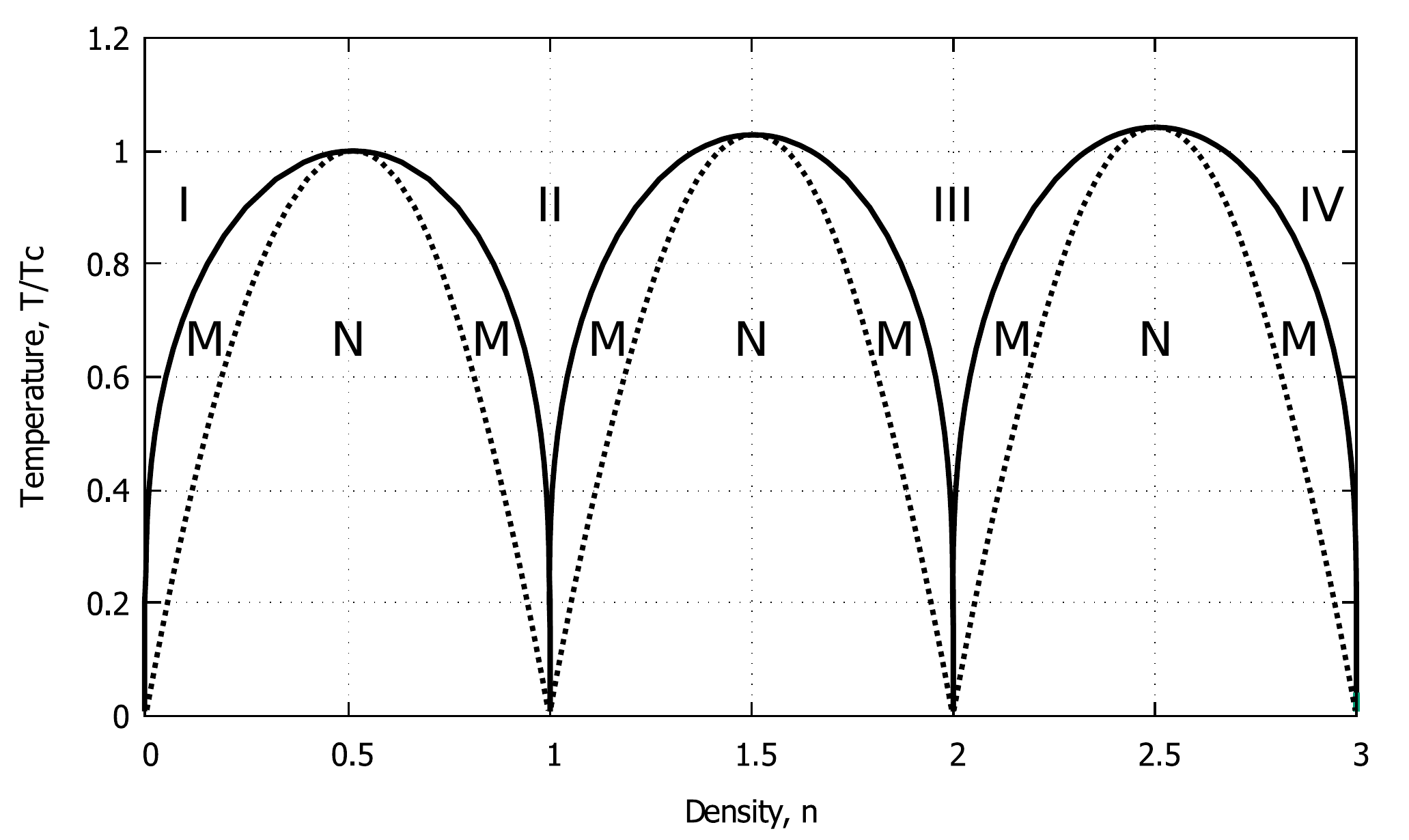}\\
	\vskip-3mm\caption{Plot of the spinodals (dotted lines) and phase coexistence curves (solid lines) between Phase I (marked by I) and Phase II (marked by II), Phase II and Phase III (marked by III), Phase III and Phase IV (marked by IV). Letter M denotes the region of metastable state, letter N \--- unstable state. ($f=1.2, v=1$).}\label{fig10}
\end{figure}
\begin{figure}[h!]
	\centering \includegraphics[width=0.45\textwidth]{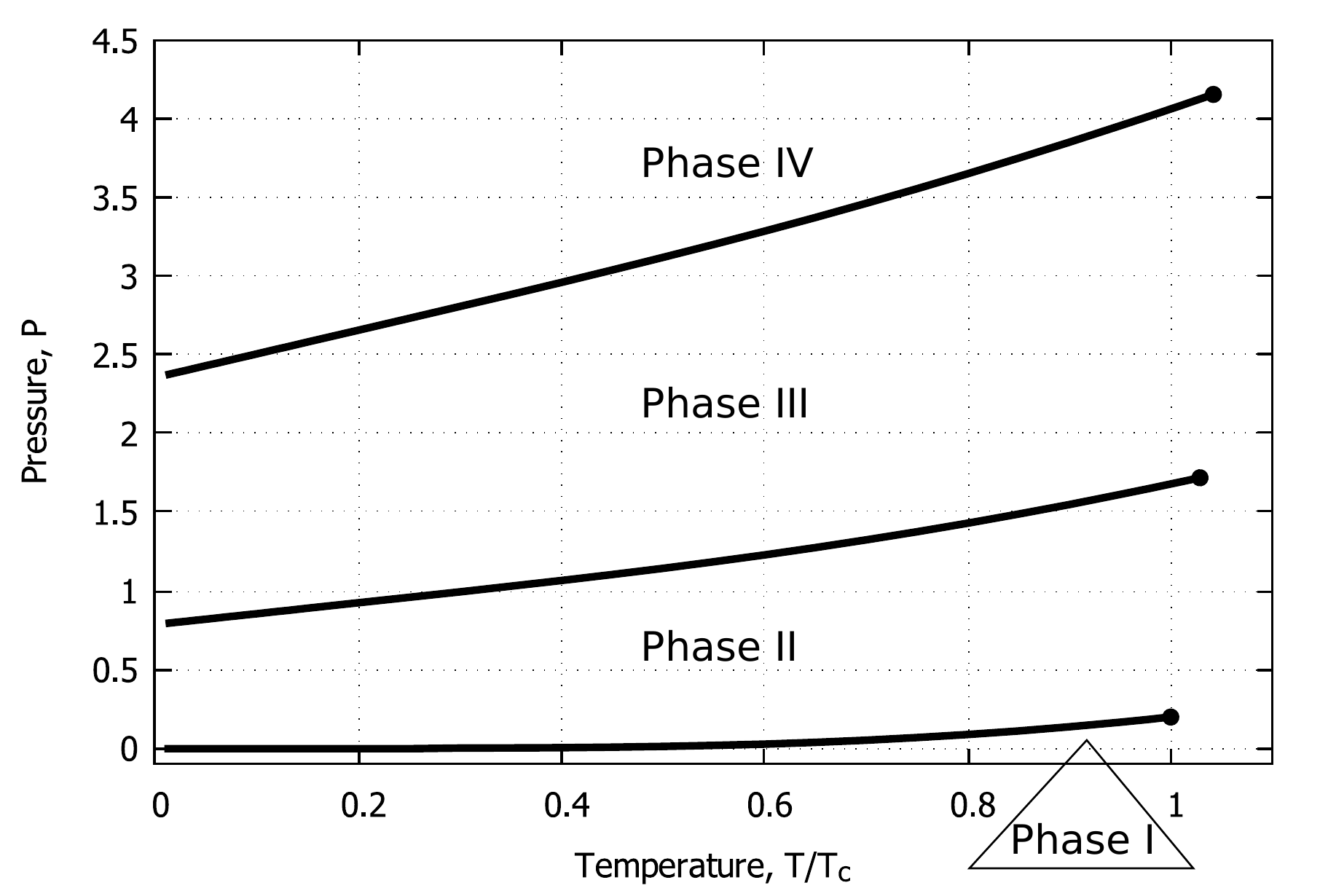}\\
	\vskip-3mm\caption{The PT phase diagram indicating first three first-order phase transitions and consistent phase coexistence between four different phases. On the right hand side each line terminates at its critical point (solid circles). ($f=1.2, v=1$).}\label{fig11}
\end{figure}

At last we present the phase diagrams in terms of $(P,\eta,T)$ (Figure~\ref{fig9}), $(\eta,T)$ (Figure~\ref{fig10}), $(P,T)$ (Figure~\ref{fig11}), based on the equation of state \eqref{4d5}. 
These figures illustrate three phase transitions in the sequence since we observe a jump of the order parameter $ \eta $ below the critical temperature (see Figure~\ref{fig9}). Thus, at the same wide temperature range, the system occurs in one of four phases. Each phase exists at its density and pressure range, and this range will be greater in value with each subsequent phase. Under certain conditions, the system can smoothly move from a stable phase to a metastable state (see Figure~\ref{fig10}), in which an unstable equilibrium occurs. The metastable region rapidly narrows as it approaches either critical points or ultra-low temperatures. In addition, the figures present that the temperature $ T_c = T_c^{(1)} $ is critical for the first-order phase transition from Phase I to Phase II. However, for subsequent phase transitions there are still density jumps at this temperature, and the corresponding $T_c < T_c^{(2)} < T_c^{(3)}$. 

The phase diagram in Figure~\ref{fig11} shows that the coexistence curves between Phase I and Phase II, Phase II and Phase III, Phase III and Phase IV slightly approach each other as the temperature decreases to zero. However, they do not intersect. This fact evidences that there is no triple point in the cell model with the Curie-Weiss interaction such that the repulsion to attraction ratio is $f=1.2$.

\renewcommand{\theequation}{\arabic{section}.\arabic{equation}}
\section{Conclusions}
\setcounter{equation}{0}

We obtained an explicit equation of state of the cell model with Curie-Weiss interaction due to the exact calculation of the grand partition function. Thus, the present research concerns elaborating the theory of the first-order phase transitions, which does not require expansions in small parameters and other approximations. We show that a sequence of phase transitions occurs in the model and propose a mathematically strict definition of the critical temperature as a function of the microscopic parameters of the model. In addition, we found the values ??of critical temperatures in the region of the first three phase transitions.  
Each of these values is larger than the previous one and corresponds to every time larger value of the critical density.

We obtained the isotherms of chemical potential as a function of density. In the temperature range $ T <T_c $ they have an increasing step-like shape. The first-order phase transition occurs at the chemical potential values corresponding to the plateau parts on the isotherms. This method allows us to avoid the Maxwell construction but instead find the points of phase coexistence without going beyond the microscopic approach.

A 3D pressure-density-temperature $(P,T,\eta)$ diagram of the cell model with Curie-Weiss interaction based on the accurate equation of state we derived demonstrates the sequence of first-order phase transitions at $ T <T_c $. The temperature-density $(T,\eta)$ diagram presents the phase coexistence curves and the spinodal, which outline the boundaries of stable phases, metastable states, and areas of instability. The coexistence curves plotted in the pressure-temperature $(P,T)$ phase diagram behave in a way that indicates the absence of a triple point in the cell model with a given set of microscopic parameters.

\renewcommand{\theequation}{\arabic{section}.\arabic{equation}}
\section{Acknowledgements}
\setcounter{equation}{0}
This work was partly supported by SRP № II-5-21, Registration number in UkrINTEI 0121U111718.


\begin{thebibliography}{99}
	\bibitem{ref1} Kozitsky Y., Kozlovskii M., Dobush O., In \emph{Modern Problems of Molecular Physics}, Springer, Cham, 2018. doi:10.1007/978-3-319-61109-9\_11
	\bibitem{ref2} Kozitsky Yu.V., Kozlovskii M.P., Dobush O.A., Condens. Mat. Phys. {\bf 23}.2 (2020) 23502. doi:10.5488/CMP.23.23502
	
	\bibitem{ref3} Kozlovskii M., Dobush O., J. Mol. Liq. {\bf 215} (2016) 58. doi:10.1016/j.molliq.2015.12.018
	
	\bibitem{ref4} Kozlovskii M.P., Pylyuk I.V., Dobush O.A., Condens. Mat. Phys. {\bf 21}.4 (2018) 43502. doi:10.5488/CMP.21.43502
	
	\bibitem{ref5} Kozlovskii M.P., Pylyuk I.V., Prytula O.O., Phys. Rev. B {\bf 73} (2006) 174406. doi:10.1103/PhysRevB.73.174406
	\bibitem{ref6} Kozlovskii M.P., Romanik R.V., J. Mol. Liq. {\bf 167} (2012) 14. doi:10.1016/j.molliq.2011.12.003
	
	\bibitem{ref12} Belotskii E.D., Lev B.I., Phys. Let. A {\bf 147}.1 (1990) 13. doi:10.1016/0375-9601(90)90005-9
	\bibitem{ref13} Lev B.I., Phys. Rev. E {\bf 58}.3 (1998) R2681. doi:10.1103/PhysRevE.58.R2681
	\bibitem{ref14} Petrenko S.M., Rebenko A.L., Tretychnyi M.V., Ukr. Math. J. {\bf 63} (2011) 425. doi:10.1007/s11253-011-0513-0
	\bibitem{ref15} Rebenko A.L., Rev. Math. Phys. {\bf 25} (2013) 1330006. doi:10.1142/S0129055X13300069
	\bibitem{ref16} Kac M., Uhlenbeck G.E., Hemmer P.C. Journ. Math. Phys. {\bf{4}}, (1963) 216. doi:10.1063/1.1703946

	\bibitem{ref17} Balesku R.C. Equilibrium and non-equilibrium statistical mechanics. Wiley. New Tork-London-Sydney-Toronto, (1978) 742.

	\bibitem{ref18} 
	Lebowitz J.L., Penrose O., J. Math. Phys. {\bf 7} (1966)  98. doi:10.1063/1.1704821

	\bibitem{ref8} Hill T. L. Statistical Mechanics: Principles and Selected Applications. Dover Books on Physics. New York : McGraw-Hill, (1956) 432.
	
	\bibitem{ref19} 
	Fedoryuk M.V., In: Analysis I: Integral Representations and Asymptotic Methods, Encyclopaedia of Mathematical Sciences, Vol.~13, Evgrafov M.A., Gamkrelidze R.V. (Eds.), Springer-Verlag Berlin Heidelberg, 1989, 83--191.

	\bibitem{ref20} 
	Van Hove L. Physica {\bf 15}.11-12 (1949) 951-961. doi:10.1016/0031-8914(49)90059-2
	
	
	
\end{thebibliography}
\end{document}